\documentclass[twocolumn]{aastex61}
\pdfoutput=1 
\usepackage{amsmath,amstext}
\usepackage[T1]{fontenc}
\usepackage{apjfonts} 
\usepackage[figure,figure*]{hypcap}
\usepackage{graphicx}
\usepackage{appendix}

\usepackage{xcolor}

\shorttitle{machine learning Abundances in NGC~1856}
\shortauthors{R.\, Asa'd, et al.}

\begin{document}

\title{NGC~1856: Using machine learning techniques to uncover detailed stellar abundances from MUSE data}
\author{Randa Asa'd}
\affiliation{American University of Sharjah, Physics Department, P.O. Box 26666, Sharjah, UAE}

\author{S. Hernandez}
\affiliation{AURA for ESA, Space Telescope Science Institute, 3700 San Martin Drive, Baltimore, MD 21218, USA}

\author{J.M John}
\affiliation{American University of Sharjah, Physics Department, P.O. Box 26666, Sharjah, UAE}

\author{M. Alfaro-Cuello}
\affiliation{Facultad de Ingenier\'{i}a y Arquitectura, Universidad Central de Chile, Av. Francisco de Aguirre 0405, La Serena, Coquimbo, Chile}
\affiliation{Space Telescope Science Institute, 3700 San Martin Drive, Baltimore, MD 21218, USA}

\author{Z. Wang}
\affiliation{Sydney Institute for Astronomy, School of Physics, A28, The University of Sydney, NSW, 2006, Australia}
\affiliation{ARC Centre of Excellence for All Sky Astrophysics in Three Dimensions (ASTRO-3D)}

\author{A. As'ad}
\affiliation{College of Wooster, Department of Mathematical and Computational Sciences, 1189 Beall Ave, Wooster, OH 44691, USA}

\author{A. Vasini}
\affiliation{Dipartimento di Fisica, Sezione di Astronomia, Università degli studi di Trieste, Via G.B. Tiepolo 11, I-34143 Trieste, Italy}

\author{F. Matteucci}
\affiliation{Dipartimento di Fisica, Sezione di Astronomia, Università degli studi di Trieste, Via G.B. Tiepolo 11, I-34143 Trieste, Italy}
\affiliation{INAF, Osservatorio Astronomico di Trieste, Via Tiepolo 11, I-34131 Trieste, Italy; INFN, Sezione di Trieste, Via Valerio 2, I.34127 Trieste, Italy}

\correspondingauthor{Randa Asa'd}
\email{raasad@aus.edu}

\begin{abstract}

We present the first application of the novel approach based on data-driven machine learning methods applied to \textit{Multi-Unit Spectroscopic Explorer} (MUSE) field data to derive stellar abundances of star clusters. MUSE has been used to target more than 10,000 fields, and it is unique in its ability to study dense stellar fields such as stellar clusters providing spectra for each individual star.
We use MUSE data of the extragalactic young stellar cluster NGC 1856, located in the Large Magellanic Cloud (LMC). We present the individual stellar [Fe/H] abundance of 327 cluster members in addition to [Mg/Fe], [Si/Fe], [Ti/Fe], [C/Fe], [Ni/Fe], and [Cr/Fe] abundances of subsample sets. Our results match the LMC abundances obtained in the literature for [Mg/Fe], [Ti/Fe], [Ni/Fe], and [Cr/Fe]. This study is the first to derive [Si/Fe] and [C/Fe] abundances for this cluster. 
The revolutionary combination of integral-field spectroscopy and data-driven modeling will allow us to understand the chemical enrichment of star clusters and their host galaxies in greater detail expanding our understanding of galaxy evolution.

\end{abstract}


\section{Introduction and Motivation}

One of the most fundamental goals of astrophysics is understanding the chemical evolution of galaxies and the universe as a whole. Only light elements formed during the Big Bang, while heavier elements were synthesized in stellar cores \citep{Alpher1948,Hoyle1954}. As a result, the type of elements and their abundances changed dramatically from the first generation of stars after the Big Bang in comparison with those been formed today. Star clusters are ideal laboratories for studying the chemical evolution of galaxies since their stars have roughly the same age, and their overall metallicity is representative of the gas reservoirs out of which the cluster stars formed. 

In addition to the overall metallicity of star clusters, abundances of individual elements that result from different nucleosynthetic processes can offer details about the time scales of the chemical enrichment. For example, a core collapse supernova would occur one billion years before the type Ia supernova. This means that the chemical enrichment with $\alpha$-elements (e.g., Ne, Mg, Si, S, Ar, Ca,  produced primarily in core collapse supernovae) is not at the same time as Fe-peak elements (e.g., Sc, V, Cr, Mn, Fe, Co, and Ni, produced primarily in type Ia supernovae) \citep{Tinsley1979, McWilliam1997}.

Regarding our understanding of star cluster formation, the multiple populations (MPs) phenomenon defined as variations in the light element abundances (e.g., He, C, N, O, Na, Al, and sometimes Mg) of individual stars in stellar clusters not expected from traditional stellar evolutionary processes \citep[e.g.,][and references therein]{Bastian18}. Knowing the accurate abundances of star clusters can help us to solve this puzzling phenomenon.

Traditionally, stellar abundances are obtained from fitting high-resolution high-signal-to-noise spectra of carefully selected individual stars within the cluster \citep{Joffre2019}. However, this method can be applied only within the Milky Way and nearby galaxies where star clusters can be resolved. At larger distances one has to rely on the abundances of brighter stars like red supergiants (RSGs) \citep{Davies10, Davies15, Gazak14, Patrick15}. Nevertheless, for star clusters beyond our local group, even the brightest stars cannot be resolved. In the past decades, high-resolution integrated spectra of star clusters played a crucial role in revealing the detailed chemical abundances of star clusters beyond our local group \citep[e.g, ][]{Larsen12, Colucci12, Larsen14}. Nowadays, thanks to the advancement of both computing facilities and observational instruments, revolutionary modern approaches like machine learning can be used to obtain accurate stellar abundances. \citet{Wang22} presented a novel approach to derive abundances for stars observed in \textit{Multi-Unit Spectroscopic Explorer} (MUSE) fields, making use of data-driven machine learning methods. They determined Galactic stellar abundances for [Fe/H], [Mg/Fe], [Si/Fe], [Ti/Fe], [C/Fe], [Ni/Fe], and [Cr/Fe] with a precision better than 0.1~dex.  

Motivated by the promising results by \citet{Wang22}, in this work we apply their approach to the extragalactic star cluster NGC~1856, located at the LMC, with an estimated age of $350\pm18$~Myr \citep{Chilingarian18}. The integrated-light abundances of NGC~1856, along with those of two other LMC clusters, were recently estimated by \citet{asad2022}. Overall, the abundances derived for this LMC stellar cluster are in good agreement with those predicted by detailed chemical evolution models tailored for this specific galaxy. Interestingly, the [Mg/Fe] abundance ratio for NGC~1856 appeared to be slightly depleted compared to those inferred for other $\alpha$-elements, such as [Ca/Fe] and [Mg/Fe]. To further investigate the chemical composition of the stars in NGC~1856, in this paper we present our analysis of the publicly available MUSE observations exploiting the recently developed abundance estimation technique by \citet{Wang22}.

This is the first study that applies the machine learning approach to obtain the stellar abundances of an extragalactic star cluster with MUSE data. In section \ref{sec:obs} we introduce the observations used in this work, in section \ref{sec:method} we describe the method used to select the cluster's members. In section \ref{sec:Abundances} we present the abundances obtained in this work. We compare our results with the literature in section \ref{sec:discussion}. Finally, a summary is provided in section \ref{sec:summary}.

\section{Observations and stellar spectra extraction}\label{sec:obs}

We use archival data observed with the wide-field mode (WFM) of MUSE \citep{Bacon2010} mounted in the UT4 of the Very Large Telescope at Paranal Observatory in Chile. The data was acquired as part of the GTO program 0102.D-0270(A) (PI: S. Dreizler). We use the data acquired on December 12$^{th}$, 2018, March 6$^{th}$, 2019, and February 3$^{rd}$, 2019. The observations are centered in NGC~1856 with four science exposures of 660 seconds each for every night.
The WFM MUSE data has a field-of-view of $59.9"\times60"$, with a wavelength coverage of the spectra between 4800 and 9300~\AA, a spatial resolution of $0.2"$~px$^{-1}$, and a mean spectral resolution of $R\sim3000$. We perform the data reduction using MUSEpack\footnote{\url{ https://musepack.readthedocs.io/en/latest/index.html}} \citep{Zeidler2019}, a python based code for the data reduction and analysis of integral field spectroscopy data, and using the ESO MUSE pipeline version 2.8.1 \citep{Weilbacher2020}.

We use PampelMuse\footnote{\url{https://pampelmuse.readthedocs.io/en/latest/index.html}} \citep{Kamann2013} to perform the single stellar spectra extraction from the MUSE data. PampelMUSE fits the PSF as a function of the wavelength of the MUSE datacube using a photometry reference catalogue to identify the stars. We use a Moffat profile \citep{Moffat1969} to model the PSF as is suggested for this data by \citet{Kamann2013}.  We use as reference the photometric catalogue of NGC 1856 by \citet{Correnti15} in the F814W filter. The photometry was performed by the authors based on \textit{Hubble Space Telescope} (\textit{HST}) data in the UVIS channel of the \textit{Wide Field Camera 3} (WFC3) as part of the programme GO-13011 (PI: T. H. Puzia). 

The modeling of the PSF allows to PampelMuse to differentiate the contribution of different sources. This tool has proven to be successful for de-blending close stellar objects, even in denser and more crowded stellar systems as globular and nuclear star clusters \citep[e.g.,][]{Alfaro-Cuello2019,Kamann2020, Martens2023,Nitchai2023}. A detailed description of the de-blending and modeling of the PSF are presented in Section 4 in \citet{Kamann2013}. PampelMuse provides a ``\textit{flag}'' for each of the extrated expectrum, where 0 is a good quality spectrum ($\mathrm{SNR}\geq10$) and a \textit{flag} equal to 1, is a spectrum where more than one source is contributing and 2 if the signal-to-noise is lower than 10. We extracted individual spectra for a total of 4880 potential NGC~1856 member candidate stars with a good quality spectrum (\textit{flag} $ = 0$). We only use stars with \textit{flag} $ = 0$ to avoid the analysis on spectra that might have more than one source contributing, or stars with low signal-to-noise that can provide large errors in the abundance estimates.

\section{Analysis} \label{sec:method}

\subsection{Obtaining Radial velocities and chemical element abundances} \label{sec:ab}

We use the method introduced in \citet{Wang22} to obtain the radial velocities and abundances from the MUSE spectra. This method uses two main tools: PampelMUSE \citep{Kamann2013} and Data-Driven Payne \citep{Ting17,Xiang2019}.  Data-Driven Payne (so-called DD-Payne) is a neural network model trained by samples of LAMOST DR5 cross-correlated with GALAH DR2 \citep{Buder2018}, and APOGEE-Payne \citep{Ting2019}. When applied to LAMOST spectra, DD-Payne can measure the effective temperature (T$_{eff}$), log~g, and and up to 16 chemical abundances (depending on the wavelength coverage of the observations) with an overall precision $<0.1$~dex and $0.2 - 0.3$~dex for Cu and Ba \citep{Xiang2019} for each star. According to \citet{Wang22}, it has been verified that despite the spectral differences in MUSE and LAMOST, applying DD-Payne on the MUSE spectra modified to have the same wavelength region and spectral resolution as LAMOST,  it is possible to measure the stellar parameters with precision better than 75~K in T$_{eff}$, 0.15~dex in log g, and lower than 0.1 dex for [Fe/H], [Mg/Fe], [Si/Fe], [Ti/Fe], [C/Fe], [Ni/Fe], and [Cr/Fe] in the parameter space of 3800 $< \mathrm{T}_{eff} <$ 7000 K, -1.5 $< \mathrm{[Fe/H]} <$ 0.5 dex.

The extracted MUSE spectra ($R\sim3000$) were degraded into LAMOST spectral resolution ($R\sim1800$), and then fitted by the DD-Payne-APOGEE to measure the radial velocity and stellar abundances of each spectrum. We obtain the parameters of 937 stars.

\subsection{Selecting Cluster Members}\label{sec:members}

\begin{figure}
\epsscale{1.2}
\plotone{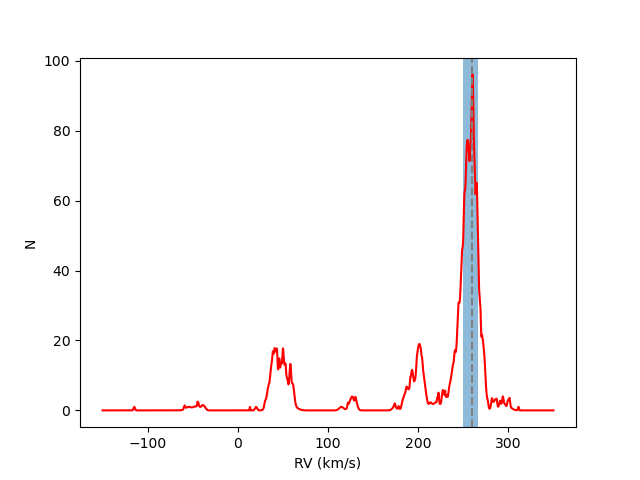}
\caption{Radial velocity distribution function for the candidate member stars of NGC~1856. The vertical dashed line indicates the mean value, and the blue shadowed region is the range of radial velocities adopted in this work. 
}
\label{RV}
\end{figure}

We perform the  selection of the cluster's members by identifying the radial velocities following the method introduced by \citet{Piatti18}. Similar to their approach, we create a distribution of radial velocities by summing the individual radial velocity values assuming a Gaussian function with a centre equal to the radial velocities of the sample selected in the previous subsection and sigma equal to their error. Figure \ref{RV} shows the results of this approach, where the shaded region represents the full width at half maximum of the primary peak. The mean, minimum, and maximum radial velocity values are 260~km~s$^{-1}$, 249.7~km~s$^{-1}$ and 266.3~km~s$^{-1}$, respectively. To ensure a reliable membership selection and exclude potential radial velocity uncertainties from the DD-Payne estimates, only stars with radial velocities within this range are considered as cluster members in our study. Our selection criteria returned a total of 467 stars classified as members of NGC~1856. \\ 

\section{Abundances Results}\label{sec:Abundances}

In this section we present the results obtained for different elements using the machine learning technique by \citet{Wang22} as detailed in section \ref{sec:ab}. For each element, the following points were taken into consideration:

\begin{enumerate}

    \item \citet{Wang22} noted that after obtaining the best overall fit value for the entire spectrum, each element has to be examined individually, since some spectral features suffer from blending due to the low spectral resolution. Their method calculates the correlation coefficient between the gradient spectra predicted by DD-Payne and the Kurucz spectral model \citep{Kurucz70, Kurucz93, Kurucz05}, to provide the measurement reliability of each element. Following the discussion about the precision of each element, we only include the results of the stars that have a correlation values higher than 0.5. 
    
    \item The method developed by \cite{Wang22} is based on specific abundance ranges for each individual element, which depends on the abundance ranges of the training set. Any values determined outside that range are estimated by extrapolation. For precision and consistency, we exclude any results obtained through extrapolation. 

\end{enumerate}

\begin{figure*}
\epsscale{0.7}
\plotone{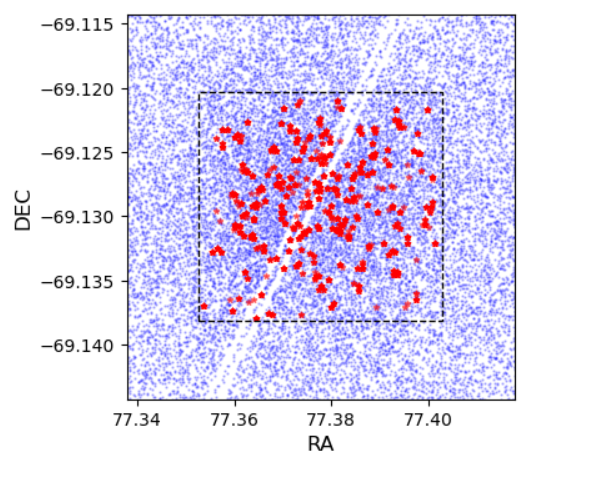}
\caption{Illustration of Cluster NGC1856: Stars in blue represent those observed by the Hubble Space Telescope (HST). The final sample (327 cluster members) highlighted in red.}
\label{cluster}
\end{figure*}

Finally, we exclude any stars with residuals (between observation spectrum and the model) $> \pm 0.2$ (20\%) and apply 2$\sigma$ clipping to further refine our selection of stars. The final sample consists of a total of 327 stars which are shown in red in Figure \ref{cluster}. 

The  cluster observed with the Hubble Space Telescope's Wide Field Camera 3 (WFC3), utilizing the UVIS (Ultraviolet and Visible) channel of the instrument and the F814W filter for optical observations under the Hubble Legacy Archive (HLA) initiative, with Proposal ID 13379 (PI Milone) is shown in blue.

\begin{figure*}
\epsscale{1.2}
\plotone{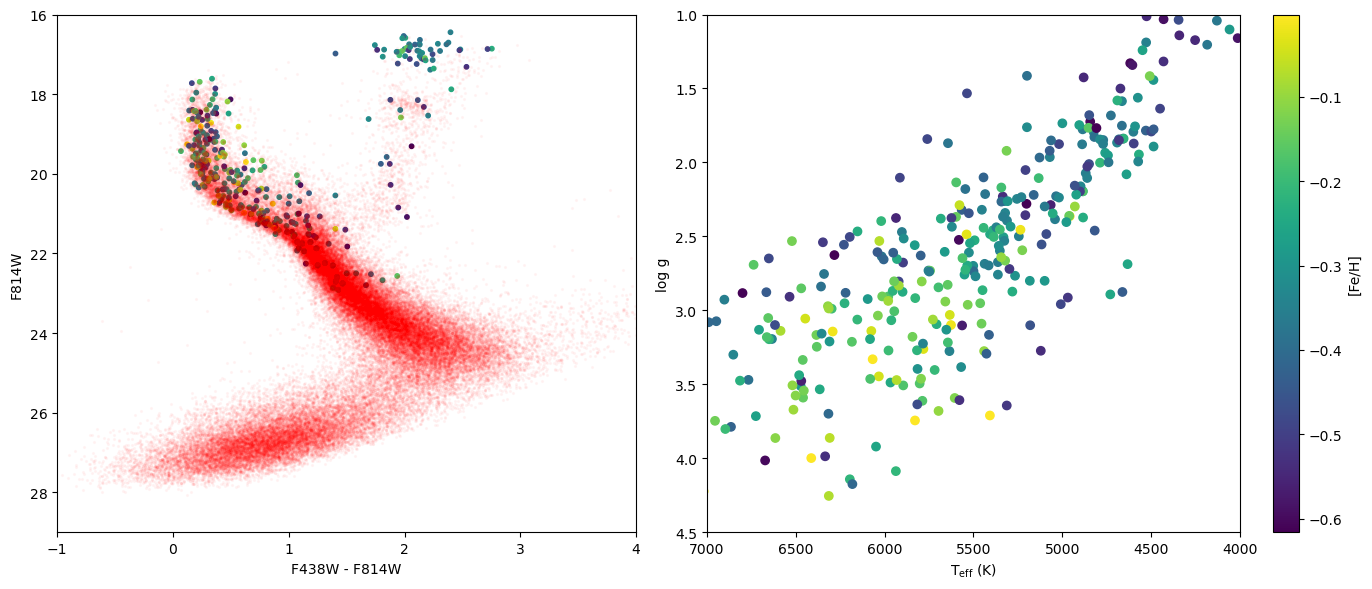}
\caption{Left panel: Color-magnitude diagram. The red points show the stars in the photometric catalogue by \citet{Correnti15}, overplotted are the stars in our final sample color-coded by metallicity. Magnitudes F438W and F418W are sourced from the photometric catalogue by \citet{Correnti15}. Right panel: The effective temperature versus log g of the stars in the final sample, color-coded by metallicity.}
\label{CMD}
\end{figure*}

Figure \ref{CMD} shows the color magnitude diagram in the left panel and the effective temperature versus log g for our final sample of 327 stars on the right panel, both are color-coded by metallicity. The residuals of our fits for these stars, along with their associated errors marked in red, are shown in the Appendix. It is a known issue that MUSE spectra underestimate the flux error (e.g., see notes of Table 1 in \citet{Alina20}) \\

Parameter errors were re-calculated by using equation 2.11 in \cite{Wang22} to represent realistic uncertainties rather than using theoretical uncertainties reported by DD-Payne. The estimated [Fe/H] values we obtained for our sample within the training set range $[-1.447,-0.500]$ has a mean  of $-0.302\pm0.154$.

\begin{figure*}
\epsscale{1.2}
\plotone{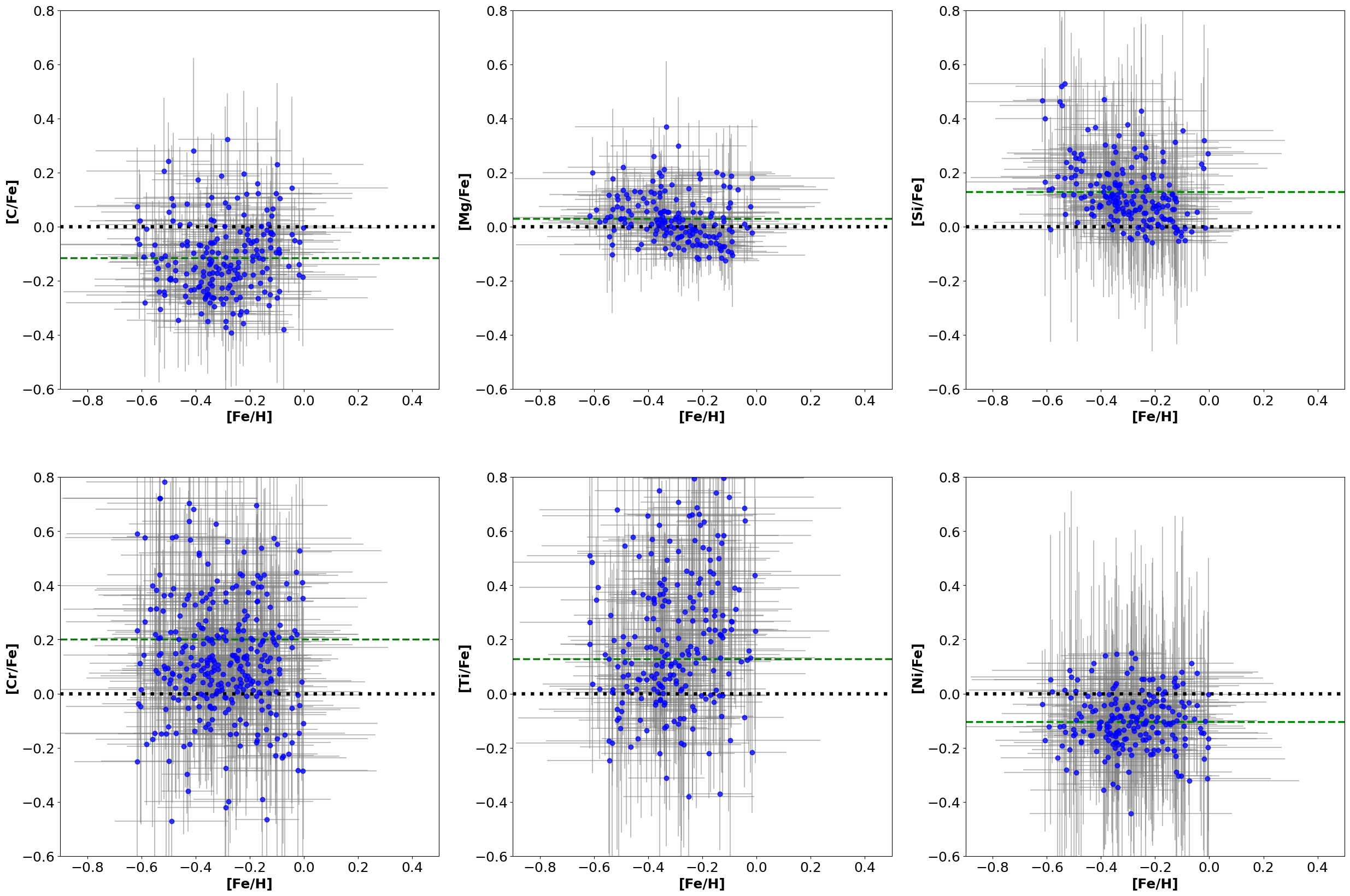}
\caption{Our inferred abundance ratios as a function of [Fe/H] excluding any results with the coefficient correlation lower than 0.5 as well as results obtained from extrapolation as explained above. The dotted line corresponds to 0.0 to guide the eye, and the dashed line represents the mean  for each element.}
\label{abundance}
\end{figure*}

In Figure \ref{abundance}, we show our inferred abundance ratios as a function of [Fe/H] excluding any results with the gradient spectra coefficient correlation lower than 0.5 as well as results obtained from extrapolation as explained above. The dotted black line corresponds to 0.0 to guide the eye, and the dashed green line represents the mean abundance value for each element. 

\begin{figure*}
\epsscale{1.2}
\plotone{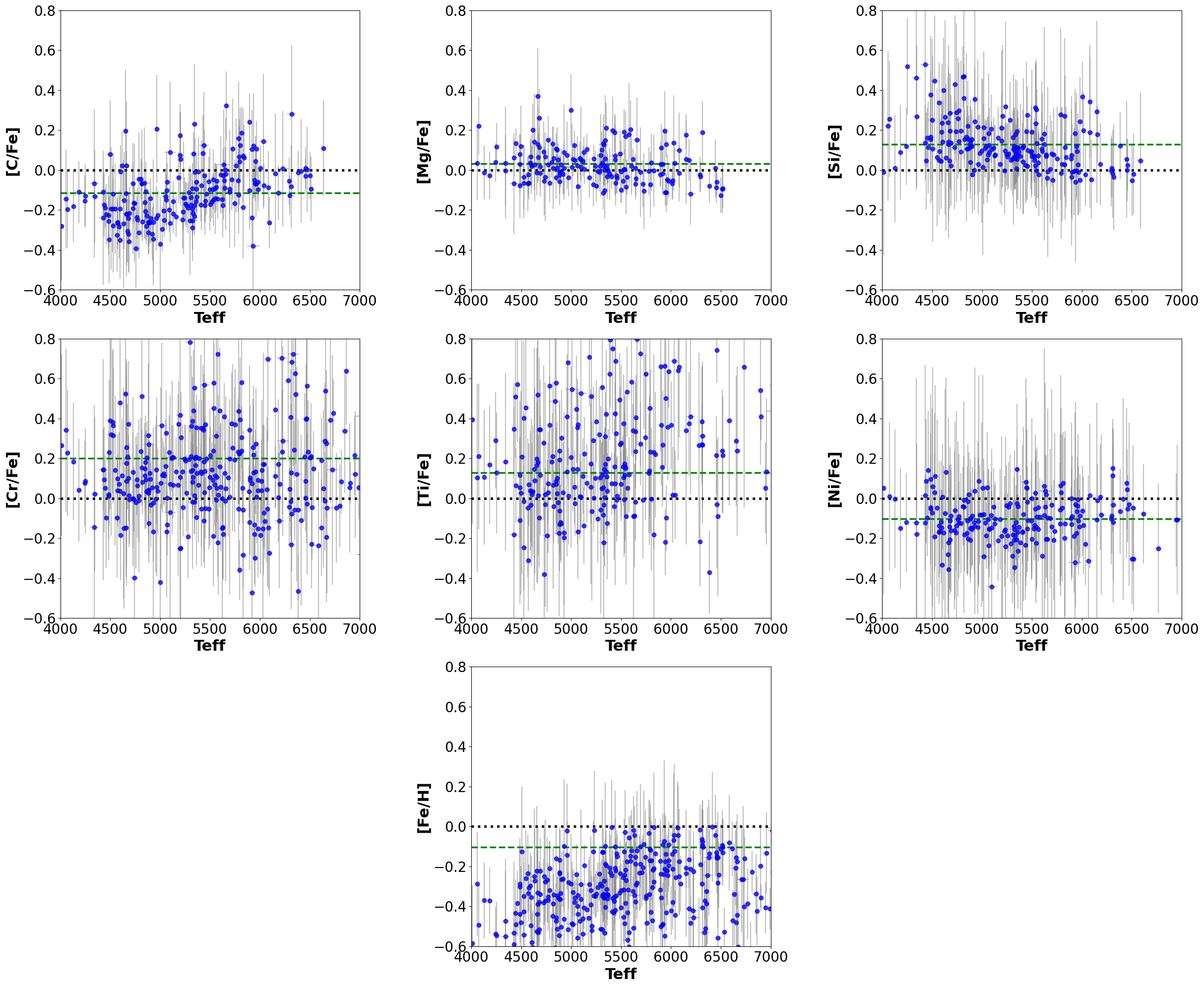}
\caption{Our inferred abundance ratios as a function of T$_{eff}$ excluding any results with the coefficient correlation lower than 0.5 as well as results obtained from extrapolation as explained above. The dotted line corresponds to 0.0 to guide the eye, and the dashed line represents the mean  for each element.}
\label{temperature}
\end{figure*}

Figure \ref{temperature} shows abundance ratios as a function of T$_{eff}$ excluding any results with the coefficient correlation lower than 0.5 as well as results obtained from extrapolation. In most cases, no trends are found. More stars are needed for a more accurate conclusion. 

\subsection{GaiaHub selection}

In order to assess our results, we repeat our analysis by using the GaiaHub pipeline \citep{delPino2022} as an additional tool for membership selection based in stellar proper motions. GaiaHub obtains proper motions by comparing \textit{Gaia} DR3 \citep{GaiaCollaboration} and \textit{HST} archival data \citep{delPino2022, Bennet22}. With GaiaHub we obtain proper motion measurements for 3,488 stars in the cluster (members and non-members). We cross-matched the GaiaHub catalogue, obtaining 539 stars in common with our single stellar spectra catalogue. From this number of stars in common, 213 are classified as members of NGC~1856 by GaiaHub (parameter \textit{"use\_for\_alignment"} as \textit{"true"}).  Analyzing the radial velocity distribution of this subset of stars, we found the mean, minimum, and maximum radial velocity values to be 264 km s$^{-1}$, 253 km s$^{-1}$, and 269 km s$^{-1}$, respectively. When adopting the selection criteria that include only stars within this range, our sample of cluster members reduced to 117 stars, further reduced to 83 after subsequent steps described in Section \ref{sec:Abundances}.\\
Despite the significant difference in the final sample sizes, we find matching abundance results within the errors for the two samples as listed in Table \ref{results}.
This shows the robustness of our analysis. This implies that the methodology can be used to even more distant and/or fainter clusters for which proper motion measurements are not available.

\section{Comparing with the literature}\label{sec:discussion}

In this section we compare our results with those obtained from thoretical models as well as other observational methods. 

\subsection{Theoretical models}

In figure \ref{literature} we present the abundance ratios as a function of [Fe/H] inferred in this work as green solid circles. The orange and light blue solid circles represent the literature data from field stars in the LMC bar and inner disc abundances, respectively, derived by \citet{VanderSwaelmen13}. The solid black line shows the chemical evolution model (Model 3 by \citet{Vasini23} - see their Table 1). Briefly, the model adopts the star formation history of \citet{Harris09} with \citet{Salpeter55} IMF. The stellar yields of \citet{Karakas10}  were adopted for the chemical enrichment from low and intermediate mass stars ($0.8 \le M/M_\odot \le 8$) and from \citet{Kobayashi06} for massive stars ($M>8M_{\odot}$). For Type Ia SNe, here assumed to originate from white dwarfs in binary systems, we adopt the yields from \citet{Iwamoto99}.
In Figure \ref{literature}, the chemical evolution model falls short in replicating key abundance patterns (Ti, C, Ni and Si). Given the recent observational confirmation of the adopted star formation history by  \cite{ruiz2020}, the likely cause of this discrepancy is probably the still uncertain yields, particularly for some of the mentioned species.To achieve a better fit to the data, we can make targeted adjustments to the stellar yields on an 'ad hoc' basis. For instance, in the case of carbon (C), we can decrease the C yields from massive stars. This adjustment is demonstrated in the lower right panel of Figure \ref{literature}, where the C yields have been reduced by a factor of 2 shown with red line. 

\cite{Nataf19} noted that the APOGEE-derived variations in [N/Fe] and [C/Fe] is consistently (at least 50\%) lower than corresponding estimates with Payne found in the literature causing a potential systematic errors within APOGEE's spectroscopic analyses. This could be the reason for the mismatch between our [C/Fe] estimates and the model. 
Concerning the stellar yields of C from massive stars, they would need to be decreased to fit the data. There are still many uncertainties in the yields calculations either due to uncertainties on nucleosynthesis or stellar evolution (mass loss and rotation). Therefore, adjustment of the yields are very ad hoc and can be used only in a heuristic way to suggest to model makers what could be done. Other element yields seem not to fit the data well and for each of them we would require  different ad hoc adjustments. We do not know a priori if a model adjusted to fit C would also fit Ti and Ni. In particular, Ni is mainly formed in Type Ia,  not much  in massive stars, and it would require adjustments for the yields of those SNe.

The nucleosynthesis of Ti element is very poorly known as noted in \cite{asad2022}. Further analysis is required to better constrain the theoretical models.

\subsection{Integrated spectra}

\begin{figure}
\epsscale{1.2}
\plotone{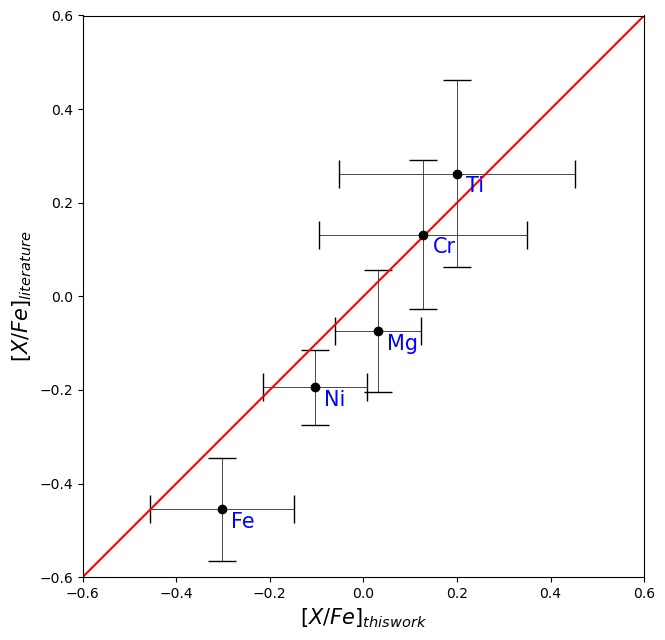}
\caption{Correlation between the mean  of the abundances obtained in this work and by \cite{asad2022} with their respective uncertainties. The red line represents the one to one relation.}
\label{compare}
\end{figure}

In order to compare our results with those derived from integrated spectra in \cite{asad2022}, we compute the mean  for each abundance along with their uncertainties obtained by dividing the standard deviation by $\sqrt{N-1}$, where $N$ is the the number of stars used for obtaining each abundance. We list in Table~\ref{results} the number of stars used in the computation of the mean  of the abundances, the mean  of the abundances obtained in this work, and the integrated abundances published by \cite{asad2022}. In Figure \ref{compare} we include a visual comparison of the abundances inferred from the integrated-light observations of NGC~1856 against the mean  values estimated after analyzing individual stars in the cluster. This figure highlights that the abundances from these different approaches are in agreement with each other within their uncertainties. 

\begin{figure*}
\epsscale{1.2}
\plotone{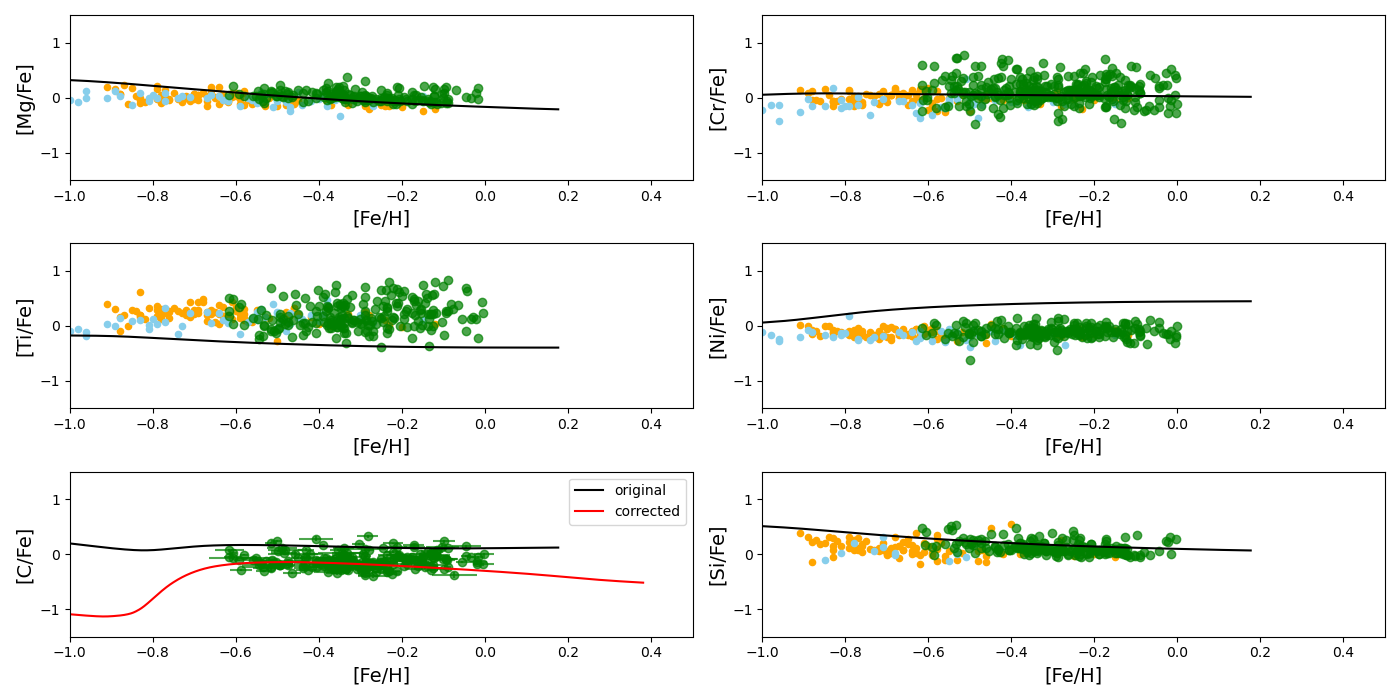}
\caption{Abundance ratios as a function of [Fe/H]. Solid green circles show the abundances inferred in this work. Solid orange and light blue circles are the literature data from field stars in the LMC bar and inner disc abundances, respectively, from \citet{VanderSwaelmen13}. The solid black line shows the chemical evolution model (Model 3 by \citep{Vasini23} (see their Table 1). In the bottom left panel, the solid red line illustrates the carbon yields from massive stars, which have been adjusted by a factor of 2 as part of an 'ad hoc' correction strategy to enhance alignment with observed data. }
\label{literature}
\end{figure*}

\cite{asad2022} found that NGC~1856 shows a possible depletion in the [Mg/Fe] abundance compared to [Ca/Fe] and [Ti/Fe]. Such a trend can be a possible hint of intracluster variations.
The top left panel of Figure \ref{literature} shows that the distribution of [Mg/Fe] is compatible with the Mg abundances of field stars, and our results in Table \ref{results} show that the [Mg/Fe] versus [Ti/Fe] values are not different considering the errors. Obtaining Na abundances of individual stars in the cluster could provide more insight into the possibility of MPs in this cluster.

\section{Summary and Future work}\label{sec:summary}

We perform the first application of the machine learning approach to obtain the stellar abundances of a extragalactic star cluster, NGC~1856, with MUSE data following the method described in \cite{Wang22}. Our main findings are summarized as follow:

\begin{itemize}

    \item We derive the first [Fe/H], [Mg/Fe], [Si/Fe], [Ti/Fe], [C/Fe], [Ni/Fe], and [Cr/Fe] abundances for individual member stars in NGC~1856.
    \item When comparing the results with those obtained when we use GaiaHub pipeline for membership selection based in stellar proper motions, we show that despite the significant difference in the final sample sizes, abundance results are matching within the errors. This implies that the technique can be applied to distant clusters for which proper motion measurements are not available.
    \item Our results are consistent with the LMC field stars abundances in the literature and the theoretical model for the abundances of [Mg/Fe], [Ti/Fe], [Ni/Fe], and [Cr/Fe].
    \item To the best knowledge, this study is the first to derive [Si/Fe] and [C/Fe] abundances for the this cluster.
    \item We find that the theoretical models of C match the observational results when we decrease the C yields from massive stars by a factor of 2.
    \item This machine learning technique is a promising tool, that can be improved by training it on MUSE-resolution observations to improve the uncertainties of the inferred values.

\end{itemize}

The work presented here highlights the potential of this technique in future studies, particularly exploiting deeper observations. This novel data-driven machine learning approach is capable of expanding our detailed abundance analysis studies of individual stars in a variety of environments beyond those observed in the Milky Way.

\begin{table}[]
\begin{tabular}{|l|cc|cc|c|}
\hline
& \multicolumn{2}{c|}{Full Sample} & \multicolumn{2}{c|} {With GaiaHub Selection} & {Literature values} \\ 
\hline
Element     & Number of Stars  & Abundance  & Number of Stars  &  Abundance & Abundance  \\
\hline
$[\text{Fe/H}]$ & 327 & $-0.302 \pm 0.15$ & 83 & $-0.339 \pm 0.10$ & $-0.455 \pm 0.11$ \\
$[\text{C/Fe}]$ & 198 & $-0.115 \pm 0.14$ & 65 & $-0.150 \pm 0.13$ & \\
$[\text{Mg/Fe}]$ & 166 & $0.031 \pm 0.09$ & 59 & $0.030 \pm 0.09$ & $-0.074 \pm 0.13$ \\
$[\text{Si/Fe}]$ & 197 & $0.129 \pm 0.12$ & 71 & $0.123 \pm 0.10$ & \\
$[\text{Ti/Fe}]$ & 220 & $0.200 \pm 0.25$ & 68 & $0.135 \pm 0.21$ & $0.262 \pm 0.2$ \\
$[\text{Cr/Fe}]$ & 311 & $0.128 \pm 0.22$ & 79 & $0.130 \pm 0.19$ & $0.132 \pm 0.16$ \\
$[\text{Ni/Fe}]$ & 205 & $-0.104 \pm 0.11$ & 65 & $-0.124 \pm 0.09$ & $-0.194 \pm 0.08$ \\
\hline
\end{tabular}
\caption{From left to right: Element ratio abundance; number of stars used in computing the 
abundance;mean of the element abundance of this work; number of stars used in computing the abundance when including the GaiaHub pipeline as an additional tool; mean of the element abundance when GaiaHub pipeline is used and the integrated abundances by \cite{asad2022}}
\label{results}
\end{table}

\section*{Acknowledgements}

We thank Paul Bennet for his help in the application of GaiaHub, Peter Zeidler for his help with MUSE data calibration and Matteo Correnti for providing the CMD data of this cluster. 
This work is supported by FRG Grant and the Open Access Program from the American University of Sharjah. This paper represents the opinions of the authors and does not mean to represent the position or opinions of the American University of Sharjah.

\clearpage
\bibliographystyle{aasjournal}
\bibliography{references}

\clearpage
\appendix 
\setcounter{figure}{0}
\renewcommand{\thefigure}{\Alph{figure}} 
\begin{figure*}[!ht]
\epsscale{1.2}
\plotone{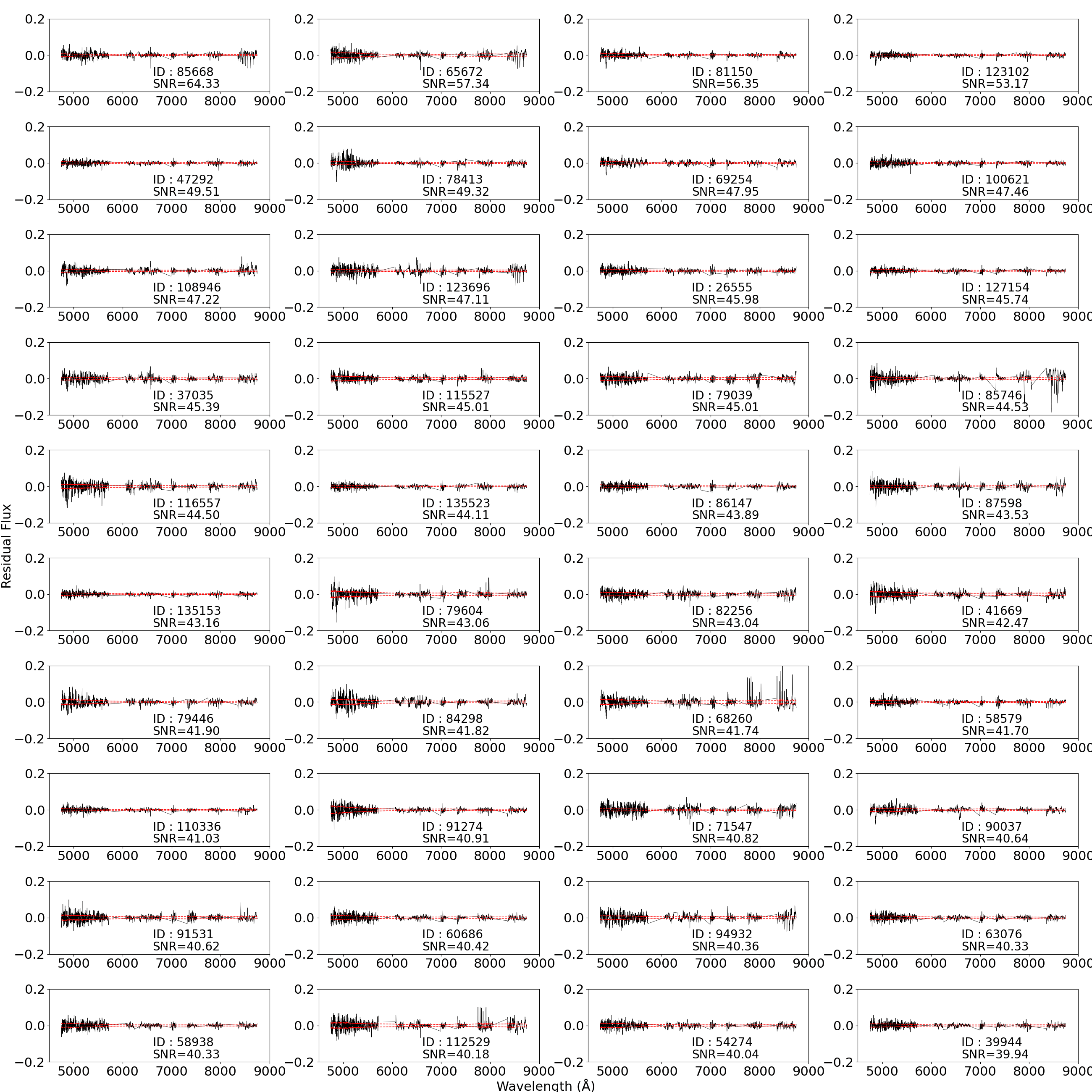}
\caption{The figure above displays the residuals of spectral fits for 40 of the final 327 stars, accompanied by the corresponding errors depicted in red.The x-axis represents wavelength in Armstrong, while the y-axis depicts flux}
\label{rich1}
\end{figure*}

\begin{figure*}
\epsscale{1.2}
\plotone{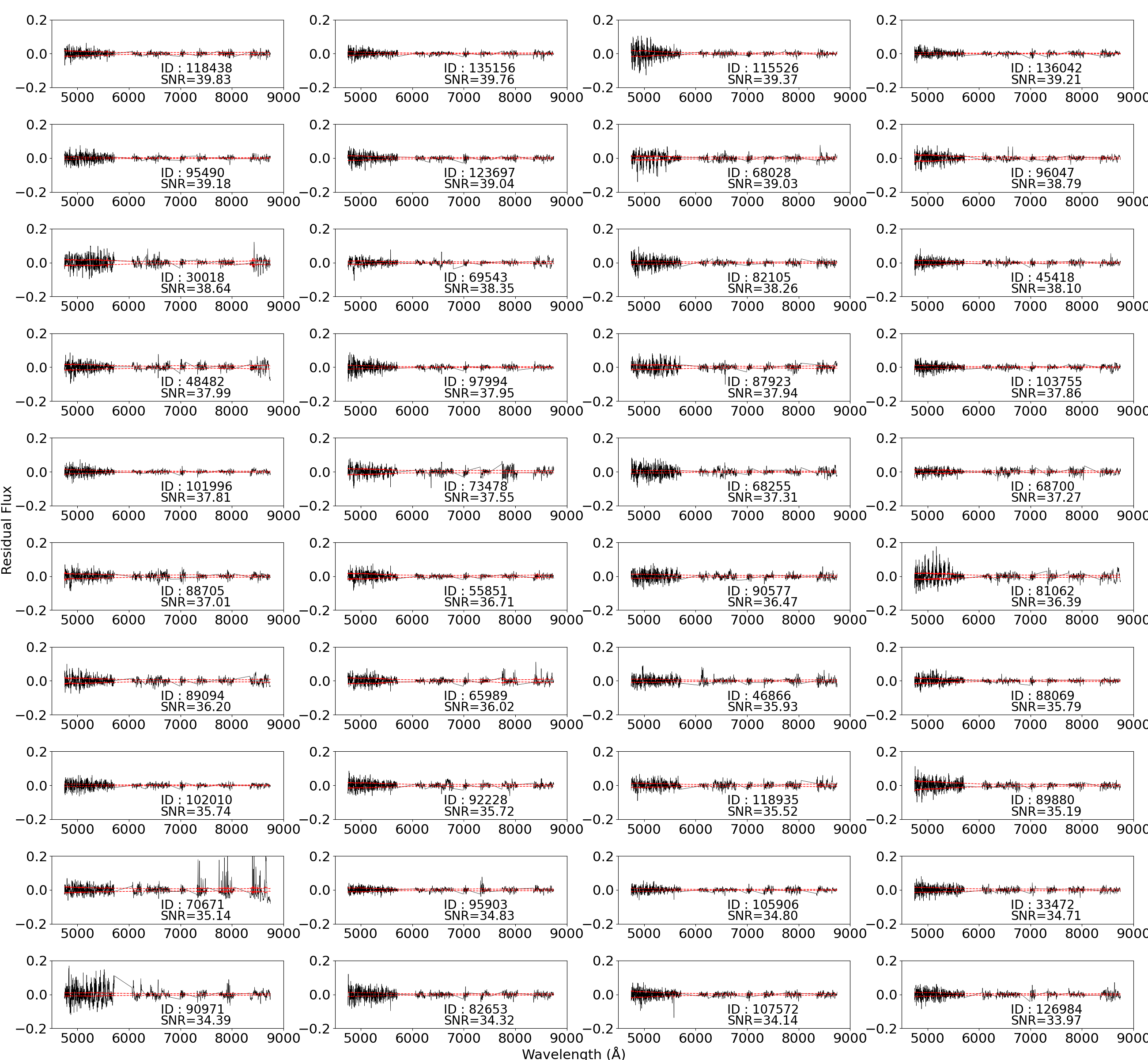}
\caption{The figure above displays the residuals of spectral fits for the second set of 40 stars from the final 327 stars, accompanied by the corresponding errors depicted in red.The x-axis represents wavelength in Armstrong, while the y-axis depicts flux}
\label{rich2}
\end{figure*}

\begin{figure*}
\epsscale{1.2}
\plotone{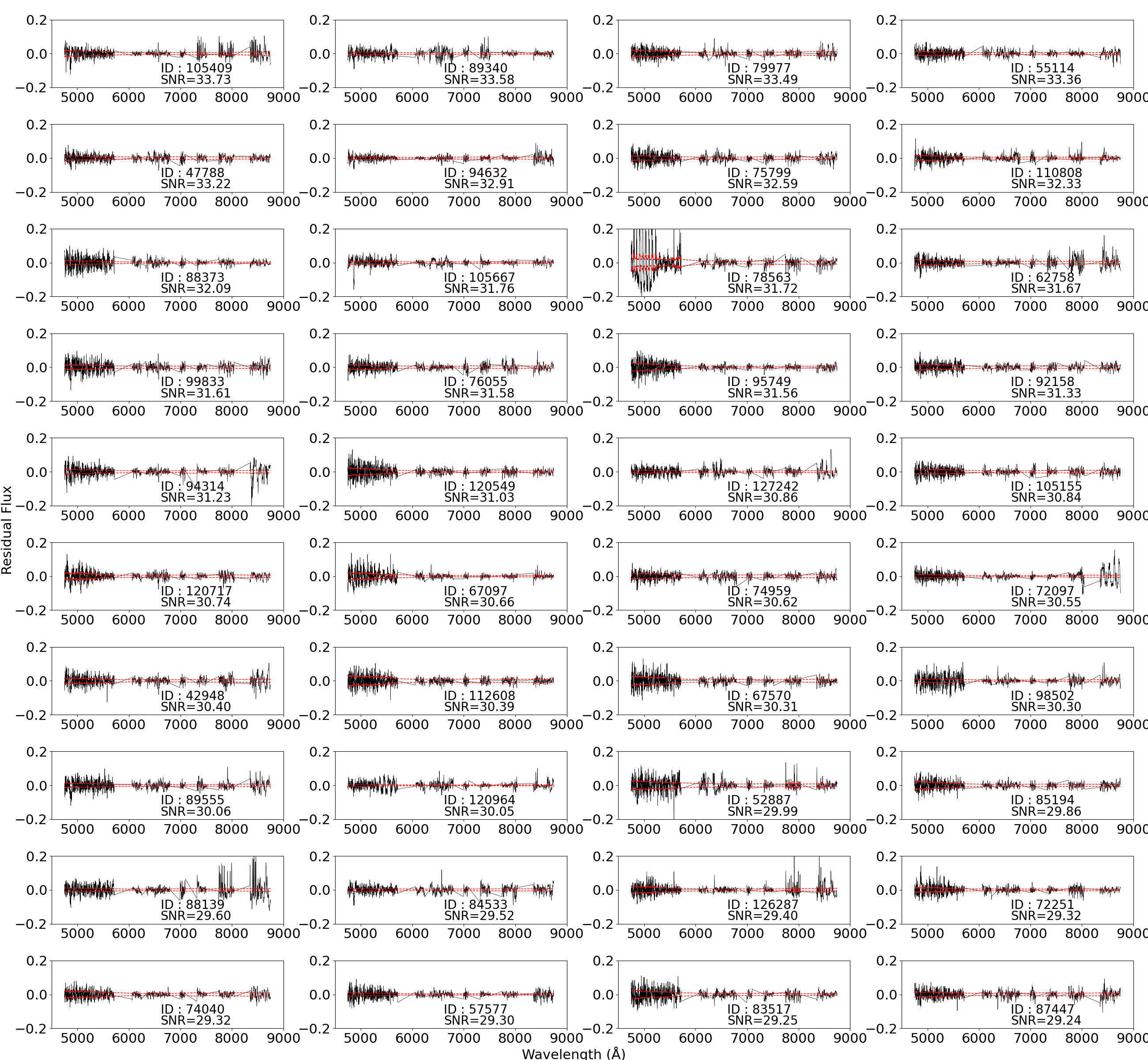}
\caption{The figure above displays the residuals of spectral fits for the third set of 40 stars from the final 327 stars, accompanied by the corresponding errors depicted in red.The x-axis represents wavelength in Armstrong, while the y-axis depicts flux}
\label{rich3}
\end{figure*}

\begin{figure*}
\epsscale{1.2}
\plotone{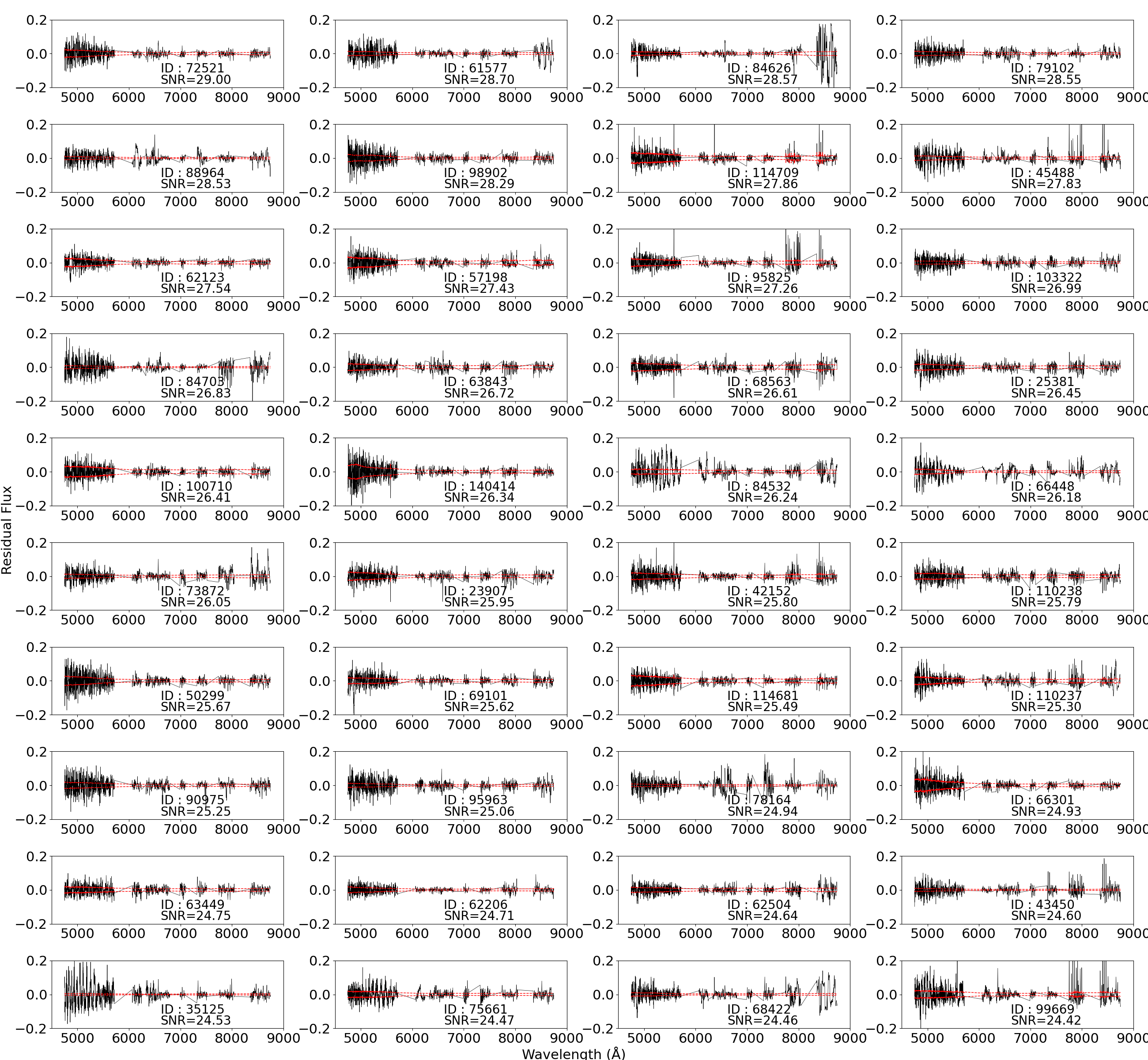}
\caption{The figure above displays the residuals of spectral fits for the fourth set of 40 stars from the final 327 stars, accompanied by the corresponding errors depicted in red.The x-axis represents wavelength in Armstrong, while the y-axis depicts flux}
\label{rich4}
\end{figure*}

\begin{figure*}
\epsscale{1.2}
\plotone{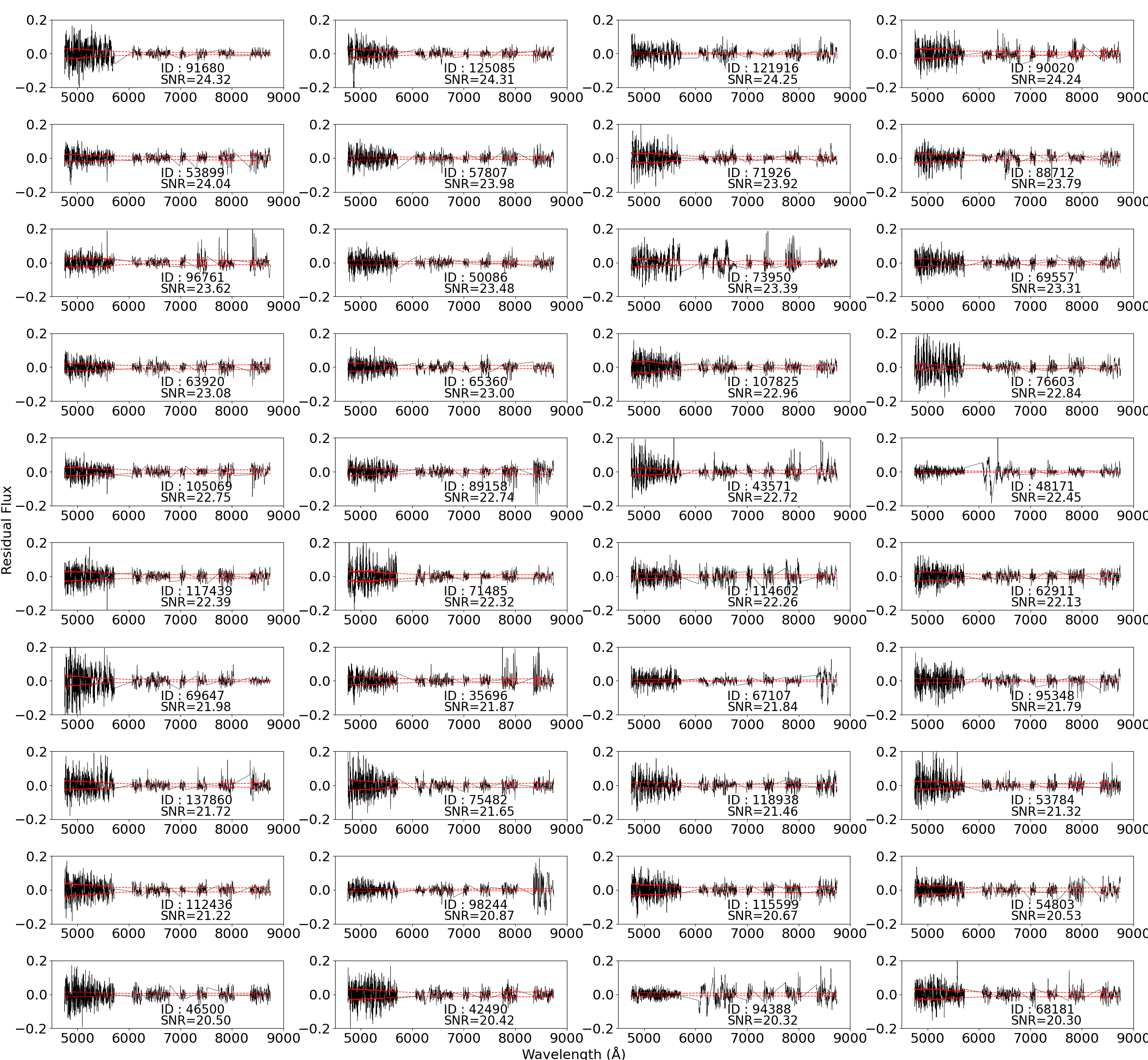}
\caption{The figure above displays the residuals of spectral fits for the fifth set of 40 stars from the final 327 stars, accompanied by the corresponding errors depicted in red.The x-axis represents wavelength in Armstrong, while the y-axis depicts flux}
\label{rich5}
\end{figure*}

\begin{figure*}
\epsscale{1.2}
\plotone{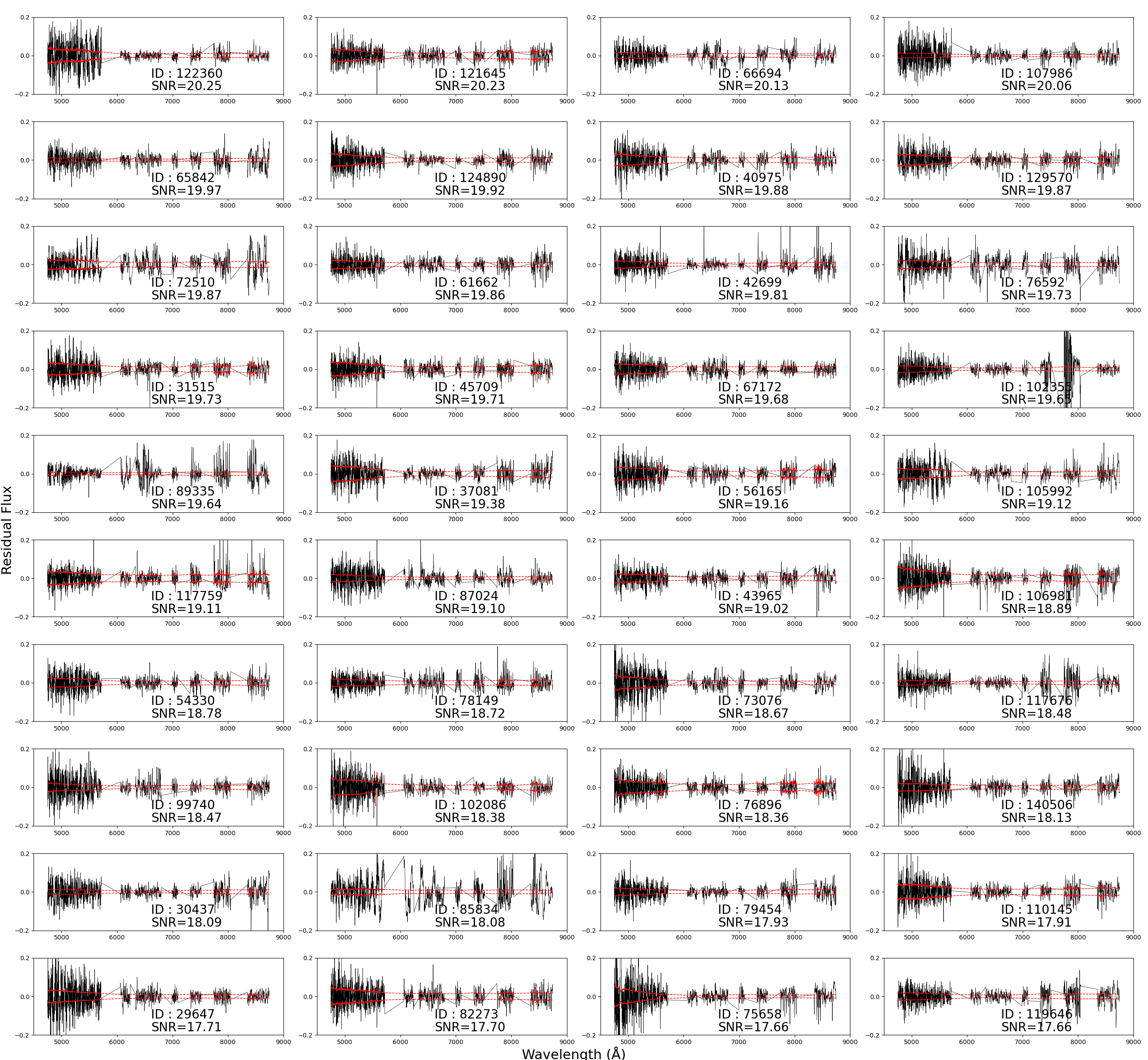}
\caption{The figure above displays the residuals of spectral fits for the sixth set of 40 stars from the final 327 stars, accompanied by the corresponding errors depicted in red.The x-axis represents wavelength in Armstrong, while the y-axis depicts flux}
\label{rich6}
\end{figure*}

\begin{figure*}
\epsscale{1.2}
\plotone{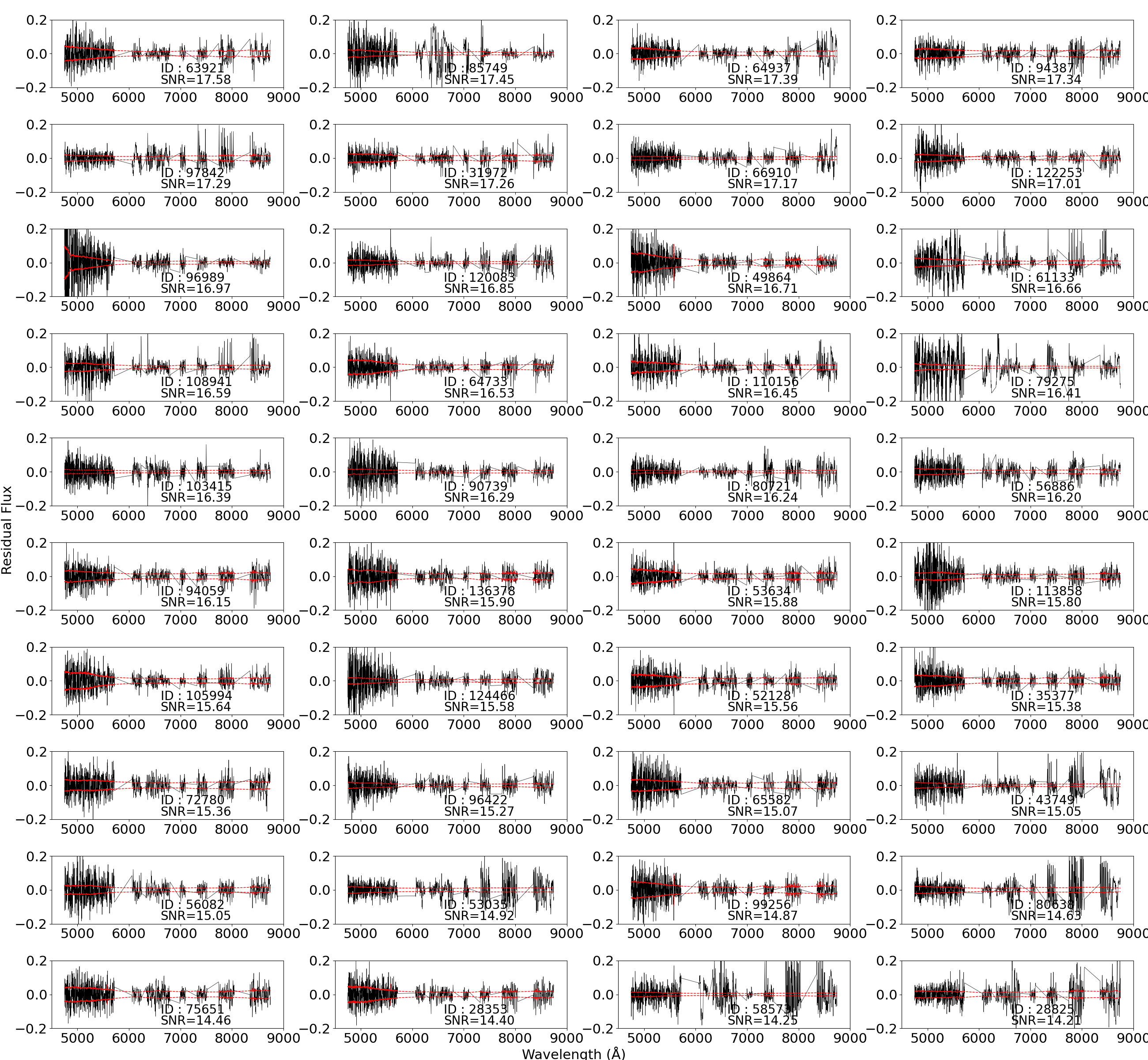}
\caption{The figure above displays the residuals of spectral fits for the seventh set of 40 stars from the final 327 stars, accompanied by the corresponding errors depicted in red.The x-axis represents wavelength in Armstrong, while the y-axis depicts flux}
\label{rich7}
\end{figure*}

\begin{figure*}
\epsscale{1.2}
\plotone{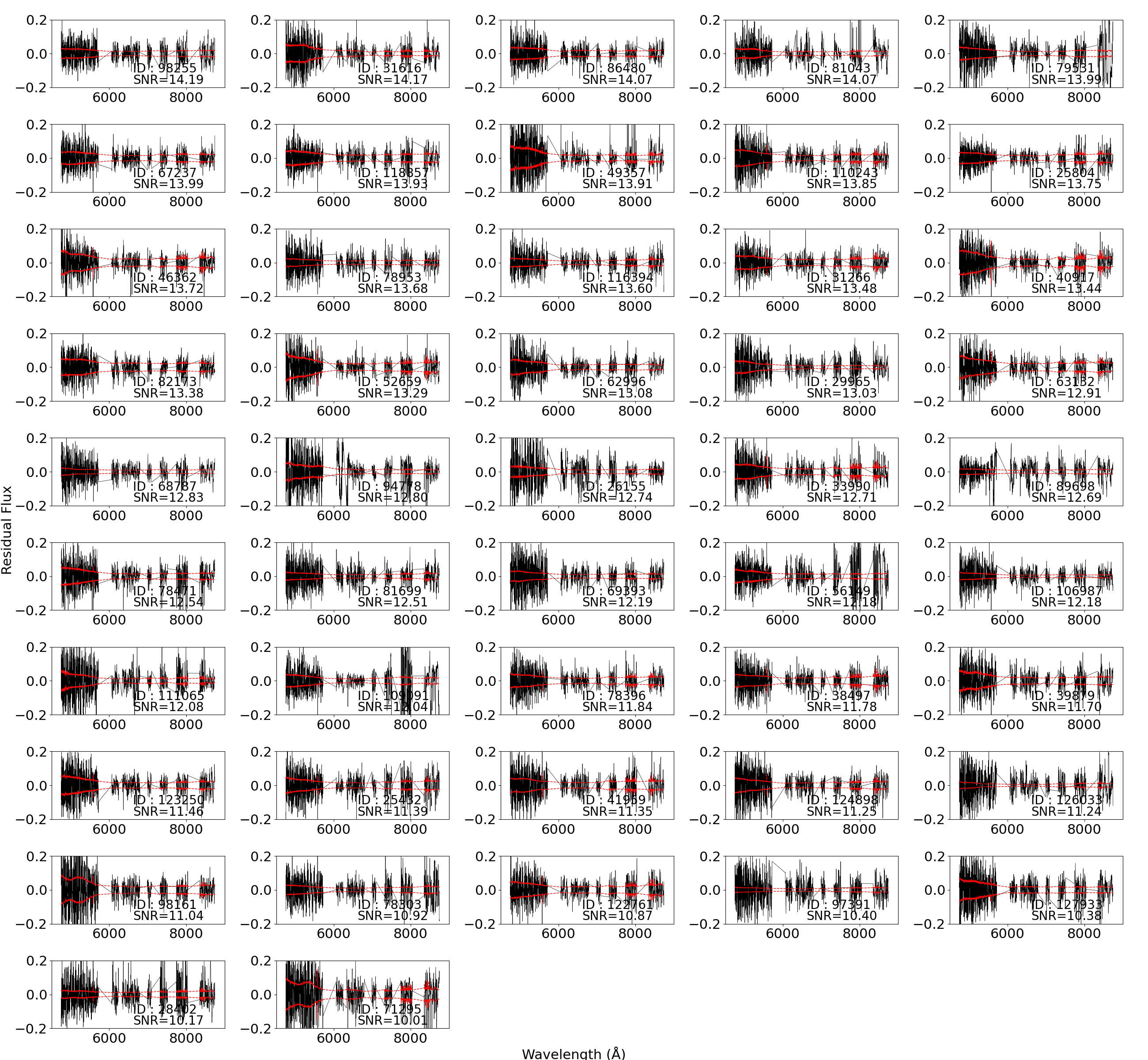}
\caption{The figure above displays the residuals of spectral fits for the last  set of 47 stars from the final 327 stars, accompanied by the corresponding errors depicted in red.The x-axis represents wavelength in Armstrong, while the y-axis depicts flux}
\label{rich8}
\end{figure*}

\end{document}